\documentclass[reprint,prb,superscriptaddress,showpacs]{revtex4-2}%
\usepackage{graphicx}
\usepackage{siunitx}
\usepackage{color}
\usepackage{amsmath}
\usepackage{amsfonts}
\usepackage{amssymb}
\usepackage{framed}
\usepackage{soul}
\usepackage{adjustbox}
\usepackage[normalem]{ulem}

\providecommand{\U}[1]{\protect\rule{.1in}{.1in}}
\makeatletter
\renewcommand*{\fnum@figure}{{\normalfont\bfseries \figurename\thefigure}}
\renewcommand*{\@caption@fignum@sep}{\textbf{ : }}
\makeatother
\usepackage{hyperref}
\hypersetup{
    colorlinks=true,
    linkcolor=blue,
    filecolor=magenta,      
    urlcolor=cyan,
}
\begin{document}

\title{Doping-induced Spin Reorientation in Kagome Magnet TmMn$_6$Sn$_6$}

\author{Mohamed El Gazzah}
\email{melgazza@nd.edu}
\affiliation{Department of Physics and Astronomy, University of Notre Dame, Notre Dame, IN 46556, USA}
\affiliation{Stravropoulos Center for Complex Quantum Matter, University of Notre Dame, Notre Dame, IN 46556, USA}
\author{Po-Hao Chang}
\affiliation{Department of Physics and Astronomy, George Mason University, Fairfax, VA 22030, USA}
\author{Y. Lee}
\affiliation{Ames Laboratory, Ames, IA, 50011, USA}
\affiliation{Quantum Science and Engineering Center, George Mason University, Fairfax, VA 22030, USA}
\author{Hari Bhandari}
\affiliation{Department of Physics and Astronomy, Rice University, Houston, Texas 77005, USA}
\affiliation{Department of Physics and Astronomy, University of Notre Dame, Notre Dame, IN 46556, USA}
\affiliation{Stravropoulos Center for Complex Quantum Matter, University of Notre Dame, Notre Dame, IN 46556, USA}
\author{Resham Regmi}
\affiliation{Department of Physics and Astronomy, University of Notre Dame, Notre Dame, IN 46556, USA}
\affiliation{Stravropoulos Center for Complex Quantum Matter, University of Notre Dame, Notre Dame, IN 46556, USA}
\author{Xiuqan Zhou}
\affiliation{Materials Science Division, Argonne National Laboratory, Lemont, IL, 60439, USA}
\author{John F. Mitchell}
\affiliation{Materials Science Division, Argonne National Laboratory, Lemont, IL, 60439, USA}
\author{Liqin Ke}
\affiliation{Ames Laboratory, Ames, IA, 50011, USA}
\author{Igor I. Mazin}
\affiliation{Department of Physics and Astronomy, George Mason University, Fairfax, VA 22030, USA}
\affiliation{Quantum Science and Engineering Center, George Mason University, Fairfax, VA 22030, USA}
\author{Nirmal J. Ghimire}
\affiliation{Department of Physics and Astronomy, University of Notre Dame, Notre Dame, IN 46556, USA}
\affiliation{Stravropoulos Center for Complex Quantum Matter, University of Notre Dame, Notre Dame, IN 46556, USA}
\date{\today}

\begin{abstract}
    The kagome-lattice compounds $R$Mn$_{6}$Sn$_6$ ($R$ is a rare earth element), where the Mn atoms form a kagome net in the basal plane, are currently attracting a great deal of attention as they have been shown to host complex magnetic textures and electronic topological states strongly sensitive to the choice of the $R$ atom. Among the magnetic $R$ atoms, TmMn$_6$Sn$_6$ orders with the easy-plane magnetization forming a complex magnetic spiral along the $c$-axis. Previous neutron studies, carried on polycrystalline, samples found that Ga doping changes the magnetic anisotropy from easy-plane to easy-axis. Here we present magnetic and magnetotransport measurements on a single crystal and first principles calculations in the doping series of TmMn$_6$Sn$_{6-x}$Ga$_x$. We find that the magnetic properties are highly sensitive even to a small concentration of Ga. With minimal Ga substitution, the easy-plane anisotropy is maintained, which gradually changes to the easy-axis anisotropy with increasing Ga. We discuss these observations with respect to the effect of Ga doping on magnetocrystalline anisotropy and Tm crystal field. 
\end{abstract} 
\maketitle



\section{Introduction}\label{sec:1}

\begin{figure}[!ht]
\begin{center}
\includegraphics[scale = 0.3 ]{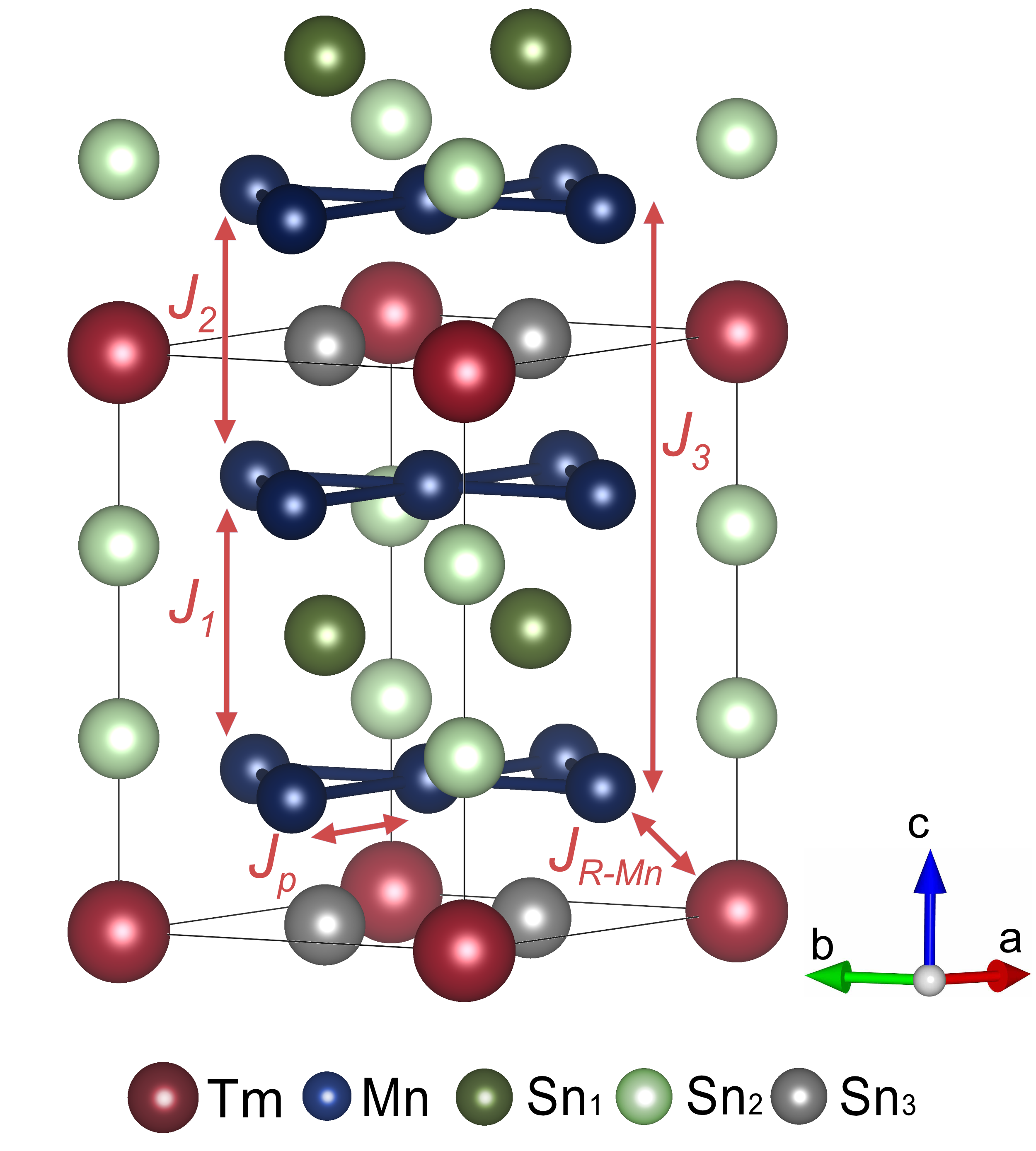}
    \caption{Schematic of the crystal structure of TmMn$_6$Sn$_6$ in the hexagonal P6/mmm space group. Sn$_1$, Sn$_2$, and Sn$_3$ are Sn atoms occupying three different inequivalent crystallographic sites  2e, 2d, and 2c, respectively. The magnetic exchange interactions between Mn between different Mn layers along the $c$-axis are denoted by $J_1$ (ferromagnetic), $J_2$ (antiferromagnetic), and $J_3$ (ferromagnetic), which play a critical role in determining the magnetic properties of the material. The in-plane exchange interaction $J_P$ is strongly ferromagnetic, and $J_{R-Mn}$ is the interaction between the Mn sublattice and the magnetic Rare-earth atoms.}
    \label{Fig1}
    \end{center}
\end{figure}

The $R$Mn$_6$Sn$_6$ family of compounds, where $R$ is a rare earth element, has recently attracted significant attention due to its unique crystal structure (shown in Fig. \ref{Fig1}), which consists of a kagome net of Mn atoms separated by two distinct slabs along the crystallographic $c$-axis \cite{ghimire_competing_2020}. This structure enables a combination of electronic features associated with the kagome lattice and a rich magnetic phase diagram arising from parametric frustration. The magnetic and electronic properties can be further tuned by the appropriate choice of the rare earth element $R$ and chemical substitutions at the sites occupied by the three constituent elements, resulting in a wide range of intriguing physical phenomena \cite{wang2021field,roychowdhury2022large,dally2021chiral,bhandari_magnetism_2024,jones_origin_2022,kitaori2021emergent,gu_robust_2022,zeng2022large,ma2021anomalous,riberolles2022prx,riberolles2024new,wenzel2022effect,samatham2024perturbation,wang2022magnetotransport,kabir2022unusual,zhang2022magnetic,jia2024nanoscale,rahman2023magnetization,madhogaria2023topological,zhu2024geometrical,bhandari2024three,destefano2025giant,fruhling2024topological}.

In compounds with non-magnetic $R$ atoms (e.g., Sc, Y, and Lu), competition among interplanar exchange interactions—$J_1$, $J_2$, and $J_3$—leads to parametric frustration, resulting in spiral magnetic ordering. The in-plane interaction between Mn moments ($J_\text{P}$) is ferromagnetic and stronger than the interplanar exchanges. In contrast, compounds with magnetic $R$ atoms exhibit an additional exchange interaction between the $R$-Sn layer and the Mn layers, denoted $J_{R\text{-Mn}}$ in Fig. \ref{Fig1}. This interaction is typically strong and dominates over the direct antiferromagnetic exchange $J_2$, thereby lifting the frustration and favoring ferrimagnetic ordering between the ferromagnetic Mn and $R$ layers \cite{jones_origin_2022}.

However, in TmMn$_6$Sn$_6$, a distinct short-pitch spiral magnetic structure is observed \cite{lefevre_neutron_2002}, indicating that the modified interaction $\tilde{J}_{2} = -2J_{R\text{-Mn}}$ is weaker than $J_2$. As a result, the net interlayer interaction remains antiferromagnetic, resembling the behavior of compounds with non-magnetic $R$ elements. In contrast, other magnetic $R$Mn$_6$Sn$_6$ compounds possess sufficiently strong $\tilde{J}_2$ to stabilize a collinear magnetic state \cite{malaman_magnetic_1999}.

Among this family, TbMn$_6$Sn$_6$ displays particularly intriguing behavior. At high temperatures, both Mn and Tb moments lie in the $ab$-plane; however, below 310 K, they spontaneously reorient toward the $c$-axis \cite{jones_origin_2022}. This spin-reorientation is driven by thermal fluctuations and marks a regime where the in-plane anisotropy of Mn becomes comparable to the out-of-plane (easy-axis) anisotropy of Tb \cite{jones_origin_2022}. This anisotropy compensation—within an otherwise strongly anisotropic system—provides favorable conditions for stabilizing spin textures such as magnetic skyrmions, which require moderate easy-axis anisotropy. Indeed, a biskyrmion texture has been observed in TbMn$_6$Sn$_6$ near the spin-reorientation transition \cite{li_discovery_2023}, establishing this region in $R166$ compounds as a promising platform for skyrmion engineering \cite{gazzah2025skyrmion}.

Other magnetic $R$Mn$_6$Sn$_6$ compounds exhibit varied magnetic behavior depending on the strength of $\tilde{J}_2$ and the on-site anisotropy of the $R$ atoms, which arises from the interplay of crystal field effects and $f$-orbital occupancy. For instance, DyMn$_6$Sn$_6$ and HoMn$_6$Sn$_6$ show canted ferrimagnetic configurations \cite{gao_anomalous_2021, min_topological_2022, guo_magnetocrystalline_2007}. TmMn$_6$Sn$_6$ is particularly interesting because it bridges the two classes: its spins form a canted arrangement like other magnetic $R166$ members, yet it also supports a short-pitch spiral state, reminiscent of non-magnetic $R166$ compounds \cite{ghimire_competing_2020, clatterbuck_magnetic_1999}.

Substituting Sn with Ga in TmMn$_6$Sn$_6$ (yielding TmMn$_6$Sn$_{6-x}$Ga$_x$) provides a tunable control parameter that modifies magnetic anisotropy and enables the stabilization of diverse magnetic phases, including a spin-reorientation transition akin to that in TbMn$_6$Sn$_6$. Therefore, a detailed investigation of the magnetic phase diagram of TmMn$_6$Sn$_{6-x}$Ga$_x$—along with an understanding of the microscopic origin of its spin reorientation—is crucial for controlling the observed topological spin textures in this system \cite{gazzah2025skyrmion}, and potentially in other $R166$ compounds.  

In this study, we investigate the magnetic properties of single-crystalline TmMn$_6$Sn$_{6-x}$Ga$_x$ samples, complemented by first-principles calculations, to explore the effects of Ga doping on magnetic anisotropy and spin reorientation. Our results reveal that at low Ga concentrations ( $0< x \lesssim 0.5$), TmMn$_6$Sn$_{6-x}$Ga$_x$ maintains an in-plane ferro/ferrimagnetic \cite{lefevre_neutron_2002} order throughout all temperatures, while higher Ga concentrations induce a clear spin reorientation transition (SRT), consistent with the previous reports \cite{lefevre_neutron_2002, canepa_magnetisation_2005}.
Upon further doping, the SRT temperature, $T_{\text{SR}}$, keeps increasing, and at $x\sim 2$ merges with the N\`eel temperature, so that the order is easy-axis at all temperatures.
Furthermore, our theoretical calculations confirm a strong preference for easy-axis magnetic anisotropy driven by Ga doping, demonstrating its potency in controlling magnetic states, hence we  provide a first-principles explanation of the curious doping dependence of $T_{\text{SR}}$. 

This work not only highlights the ability to manipulate anisotropy to achieve targeted magnetic phases but also lays a foundational understanding for the future exploration of manipulating magnetic anisotropy in the $R166$ family and beyond.

\section{Methods}
\subsection{Experimental Methods}\label{sec:2}

\textbf{Single crystal growth.} Single crystals of TmMn$_6$Sn$_{6-x}$Ga$_x$ were grown by the self-flux  method using Sn as a flux. Tm pieces (Alfa Aesar; 99.9\%), Mn pieces (Alfa Aesar; 99.95\%), Sn shots (Alfa Aesar; 99.999\%),and Ga pieces (Alfa Aesar; X\%) were loaded into a 2 mL aluminum oxide crucible in a molar ratio of $1:6:20-x:x$ respectively, and sealed in a fused silica ampule under vacuum. The sealed ampule was heated to 1150$^{\circ}$C over 10 hours, kept at 1150$^{\circ}$C for 10 hours, and then cooled to 600$^{\circ}$C at the rate of 5$^{\circ}$C/h. Once the furnace reached 600$^{\circ}$C, the tube was centrifuged to separate the crystals in the crucible from the molten flux. Several well-faceted hexagonal crystals up to 100 mg were obtained in the crucible. We found that the nominal $x$ = 0, 0.5, 1.5, 1.8, 2.4, 2.8, 3, 4, 4.5, and 5 yielded the identified compositions of $x$ = 0, 0.2, 0.5, 0.8, 1, 1.2, 1.4, 1.7, 1.8, and 1.9, respectively. The actual compositions were inferred from the combination of energy dispersive X-ray spectroscopy (EDS) and Rietveld refinement of the powder X-ray diffraction (PXRD) pattern.  \newline

\textbf{Structural and composition characterization.} 

The crystal structure of each composition was confirmed through Rietveld refinement of powder X-ray diffraction (PXRD) data collected at room temperature using a Rigaku MiniFlex diffractometer. For each batch, a representative single crystal was cleaned to remove residual surface flux and then ground into a fine powder for diffraction measurements. Rietveld refinements were performed using the FULLPROF software package \cite{Rodriguez-carvajal1993}. Elemental composition was assessed on polished crystal surfaces using energy-dispersive X-ray spectroscopy (EDS) with an Octane Elect Plus system integrated into a JEOL JSM-IT500HRLV scanning electron microscope (SEM). Elemental maps were acquired at an accelerating voltage of 15 kV to ensure sufficient spatial resolution and signal quality.\newline

\textbf{Transport and Magnetic Measurements.} Magnetic susceptibility and magnetization measurements were carried out using a Quantum Design Dynacool Physical Property Measurement System (PPMS) with a 9-T magnet. The AC Measurement System (ACMS) option was used for these measurements. Orientation of each crystal was checked by measuring the (001) peak from the surface of the crystal using the Rigaku Miniflex X-ray diffractometer.

For electrical transport measurements samples were polished to dimensions of $~1.00\times0.40\times0.15$ mm with the long axis corresponding to a random in-plane direction while the field is applied out-of-plane. An excitation current of $2.5$ mA was used in all measurements presented. Electrical contacts were affixed using Epotek H20E silver epoxy and $25$ $\mu$m Pt wires with typical contacts resistances of $\approx 20\Omega$. MR and Hall effect data were collected using the resistivity option within a Quantum Design DynaCool PPMS equipped with a 9 T magnet. MR is defined as $MR = (\rho(B)-\rho(0))/\rho(0)$, where $\rho(B)$ and $\rho(0)$ are the longitudinal resistivity with and without an applied magnetic field, $B$. The Hall resistivity was asymetrized from the positive and negative applied magnetic fields via $\rho_{\text{H}}=[\rho_{T}(+B)-\rho_{T}(-B)]/2$, where $\rho_{T}$ is the resistivity measured via the transverse voltage contacts in the Hall bar geometry. 

\begin{figure}
\begin{center}
\includegraphics[width=\linewidth]{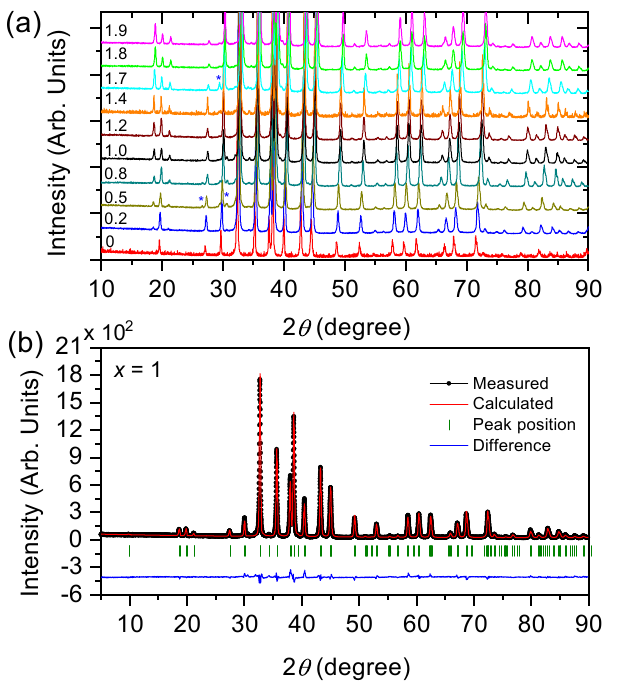}
    \caption{X-ray powder diffraction patterns and analysis of Ga-doped TmMn$_6$Sn$_6$ compounds. (a) X-ray diffraction patterns for different Ga compositions, showing the retention of the hexagonal structure across the doping series. (b) Rietveld refinement of the XRD pattern for $x$ = 1 Ga concentration, with measured and calculated patterns compared, confirming the substitution of Ga for Sn 2c site.} 
    \label{Characterization}
\end{center}
\end{figure}

\subsection{Computational Details}
DFT calculations for the total anisotropy energy are performed using a full-potential linear augmented plane wave (FP-LAPW) method, as implemented in \textsc{Wien2k} ~\cite{WIEN2k}.
The generalized gradient approximation of Perdew, Burke, and Ernzerhof ~\cite{perdew1996} is used for the correlation and exchange potentials.
To generate the self-consistent potential and charge, we employ $R_\text{MT} \cdot K_\text{max}$ = 8 with muffin-tin (MT) radii = 2.7, 2.4, 2.4, and 2.5 a.u., for Tm, Mn, Ga, and Sn atoms, respectively.
The calculations are performed with 264 irreducible $k$-points.

We employ both the DFT+$U$ method, using the fully-localized-limit (FLL) double-counting scheme, and the open-core approach to treat the strongly-correlated $R$-$4f$ electrons.
These two methods give almost identical band structures near $E_\text{F}$ and spin magnetic moments of non-$4f$ components, but they also allow us to distinguish the anisotropy contributions from the Tm-$4f$ sublattice and others~\cite{lee2023prb}.
In DFT+$U$ calculations, we use a relatively large $U$ value ($\sim 10$ eV with $J = 0$ eV) and control the initial orbital occupancy of $4f$ states to ensure that the self-consistent $4f$ electron configuration satisfies Hund's rules~\cite{lee2023prb,lee2024a} and thus consistent with the experiment.
On the other hand,  in open-core calculations, the occupied $4f$ electrons are treated as isotropic spherical charges, which allows us to calculate the anisotropy contributions from non-$4f$ electrons only.

The lattice constants used in calculations for TmMn$_{6}$Sn$_{6}$ and TmMn$_{6}$Sn$_{4}$Ga$_{2}$ were obtained from experimental TmMn$_{6}$Ga$_{x}$Sn$_{6-x}$ values with $x=0$ and 1.9, respectively.


To simulate high-temperature conditions, a more detailed analysis on non-$4f$ contributions to MA is conducted using open-core approximation implemented in VASP for thullium.

In $R166$ compounds, the Mn-$3d$ moments are strongly ferromagnetically ordered within the same layer while coupled weakly between different layers. This allows us to treat each Mn layer as a single classical magnetic moment and only focus on the interactions between layers.
We then follow the previous studies by constructing an effective mean-field 1D Heisenberg chain model \cite{jones_origin_2022,ghimire_competing_2020,ROSENFELD20081898}. A $1\times1\times2$
supercell that consists of four Mn planes is used to provide enough
configurations for extracting relevant exchange and anisotropy
parameters. In this effective chain model, three Mn sites within each layer are treated as an effective moment. We consider five distinct magnetic orderings: $uuuu$,
$uuud$, $udud$, $uudd$, and $uddu$, with their spin quantization axis (SQA) 
oriented along $z$ and $x$ axes. Here $u$($d)$ represents spin up(down) in each
Mn plane. The parameters are obtained by fitting the DFT results to
Eq. \ref{eq:spin_ham}
\begin{align}
H=E_{0}+\sum_{i<j}J_{ij}\mathbf{m}_{i}\mathbf{m}_{j}+A\sum_{i}(m_{i}^{z})^{2}+\sum_{i<j}B_{ij}m_{i}^{\mathit{z}}m_{j}^{\mathit{z}}\label{eq:spin_ham}
\end{align}
where $H$ is the total magnetic energy per Mn plane, and $m$, $E_{0}$, $J_{i,j}$, $A$ and $B_{i,j}$ correspond to the normalized effective moments ($\mathbf{m}=\mathbf{S}/|S|$ and  $|\mathbf{m}|=1$),
the reference energy, isotropic exchange, single-site anisotropy and
Ising-type anisotropic exchange, respectively. Both $J_{i,j}$ and
$B_{i,j}$ only connects a pair of adjacent Mn planes defined as $J_{i,j}=J_{n}$
and $B_{i,j}=B_{n}$ where $n=2$ if the exchange interaction pathway crosses the layer that contains the rare earth elements and otherwise $n=1$ as shown in Fig. \ref{Fig1}.  Please note that, in our definition,  we used a normalized spin variable and 
the amplitude of the Mn magnetic moment is absorbed
in the magnetic exchange parameters $J_{i,j}$, $A$ and $B_{i,j}$.

\begin{figure}[!hb]
\begin{center}
\includegraphics[width=0.8\linewidth]{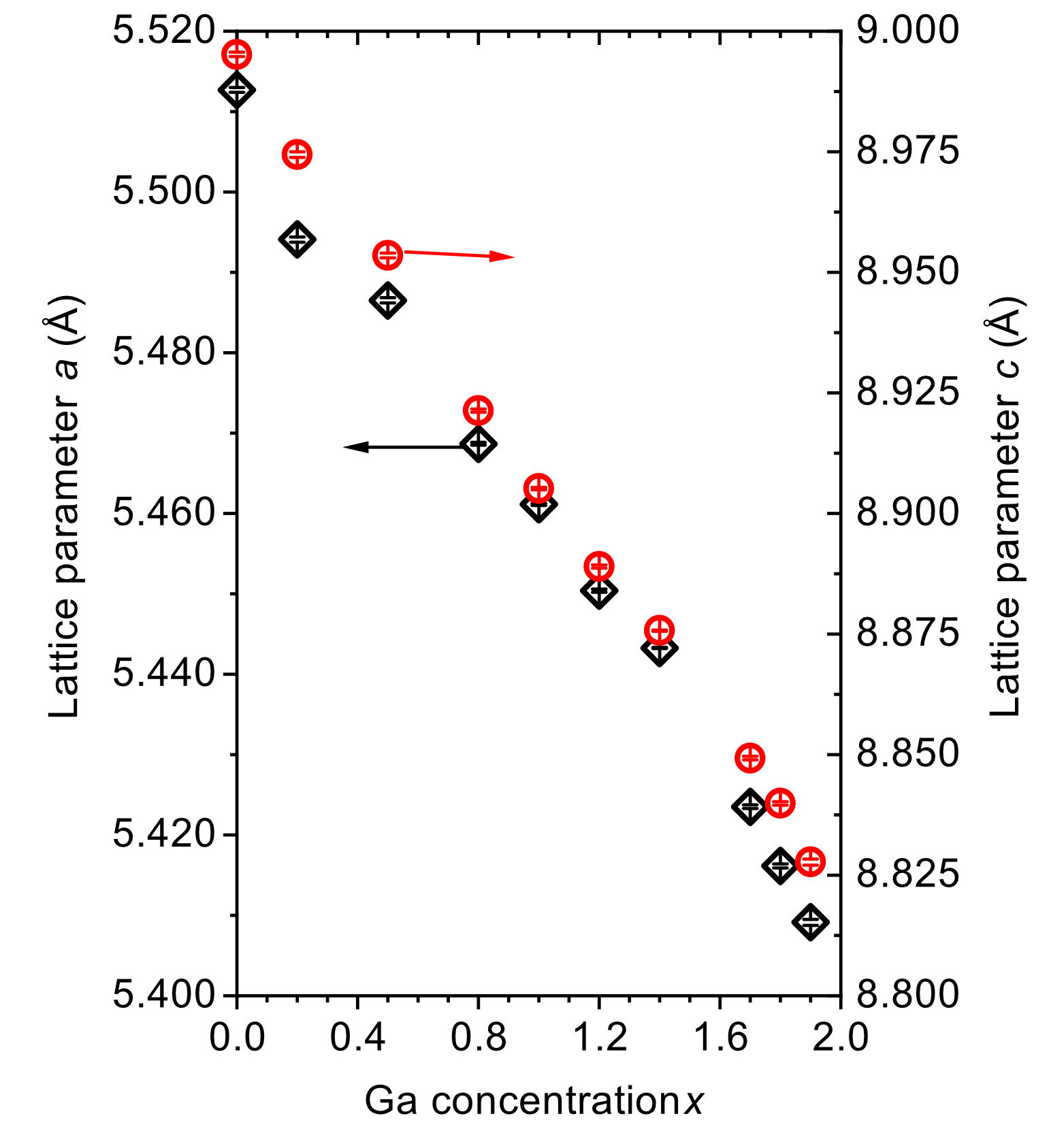}
    \caption{Lattice parameters of TmMn$_6$Sn$_{6-x}$Ga$_x$ as a function of Ga concentration $x$ determined from Rietveld refinement of PXRD collected at room temperature . The lattice parameters $a$ and $b$ (left $y$-axis) and $c$ (right $y$-axis) systematically decrease with increasing Ga content, reflecting the substitution of smaller Ga atoms for Sn in the crystal lattice.}
    \label{LatticeParameters}
    \end{center}
\end{figure}

\section{Results and Discussion}\label{sec:4}

\begin{table}[h]
 
\renewcommand{\arraystretch}{1}
\caption{Selected data from Rietveld refinement of powder
X-ray diffraction collected on ground crystals of TmMn$_6$Sn$_5$Ga$_1$.}
\begin{tabular}
 [c]{l@{\hspace{2 cm}}l@{\hspace{2 cm}}l}\hline   
Crystal system & Hexagonal \\
Space group & $P6$/mmm  \\

$a$, $b$ (\AA) & 5.45419(22)\\
$c$ (\AA)   & 8.8935(37)      \\

$R_{WP}$&$11.4\%$ \\
$R_B$ & $5.16\%$ \\
$R_F$ & $5.25\%$ \\
\end{tabular}

\par%

\begin{tabular} 
[c]{c@{\hspace{0.4cm}}c@{\hspace{0.4cm}}c@{\hspace{0.4cm}}c@{\hspace{0.4cm}}c@{\hspace{0.4cm}}c@{\hspace{0.4cm}}c}\hline
 
Atom &Wyck & x & y & z &Occupancy    \\ \hline
Tm   &   1a  &   0   & 0      &   0 & 1 \\
Mn  &  6i   & 0     & 1/2    &  0.24379 & 1 \\ 
Sn$_1$  &  2e  &  0   &  0   &   0.33520 & 1 \\
Sn$_2$   &  2d   &   1/3   & 2/3    &   1/2 & 1\\
Sn$_3$  &  2c   & 1/3    &  2/3  &  0 & 0.5\\ 
Ga  &  2c   & 1/3     & 2/3    &  0 & 0.5\\

\hline

\end{tabular}
\label{T1}
\end{table}

\subsection{Structural Characterization}

PXRD data for TmMn$_6$Sn$_{6-x}$Ga$_x$ with varying Ga concentrations confirm that the hexagonal $P6/mmm$ structure is preserved across the doping series (Figs. \ref{Characterization}(a,b)). Rietveld refinement indicates that Ga preferentially substitutes at the Sn3 site (Wyckoff position 2c), as summarized in Table \ref{T1}. Refinements were performed by systematically testing Ga occupancy across all three Sn sites (2c, 2d, and 2e). The best fit to the experimental data was obtained when Ga was restricted to the 2c site, with occupancies guided by Sn/Ga atomic ratios extracted from energy-dispersive X-ray spectroscopy (EDS). A representative refinement for the $x=1$ composition is shown in Fig. \ref{Characterization}(b).

This experimental conclusion is reaffirmed by density functional theory calculations of formation energies for TmMn$_6$Ga$_2$Sn$_4$ with Ga fully occupying each of the three inequivalent Sn sites. The Sn(2c) site yields the lowest formation energy, while substitution at Sn(2d) or Sn(2e) increases the energy by 233 meV/Ga and 547 meV/Ga, respectively. These results, consistent with prior neutron diffraction studies \cite{lefevre_neutron_2002}, confirm that Ga substitution at the 2c site is energetically favored. Accordingly, all theoretical calculations presented in this work focus on Ga occupying the Sn(2c) position.

Lattice parameters $a$, $b$, and $c$, extracted from PXRD refinements, are plotted in Fig. \ref{LatticeParameters} as a function of Ga concentration. We observe a clear trend where $a$, $b$ (left y-axis), and $c$ (right y-axis) decrease with increasing Ga concentration. This contraction is consistent with the smaller ionic radius of Ga relative to Sn and provides further confirmation of successful substitution. Such structural modifications directly impact the magnetic exchange interactions and, subsequently, the magnetic anisotropy and spin dynamics of the system.

\subsection{Magnetic Anisotropy and Spin Reorientation Induced by Ga Doping}

\begin{figure}[!ht]
\begin{center}
\includegraphics[scale=0.7]{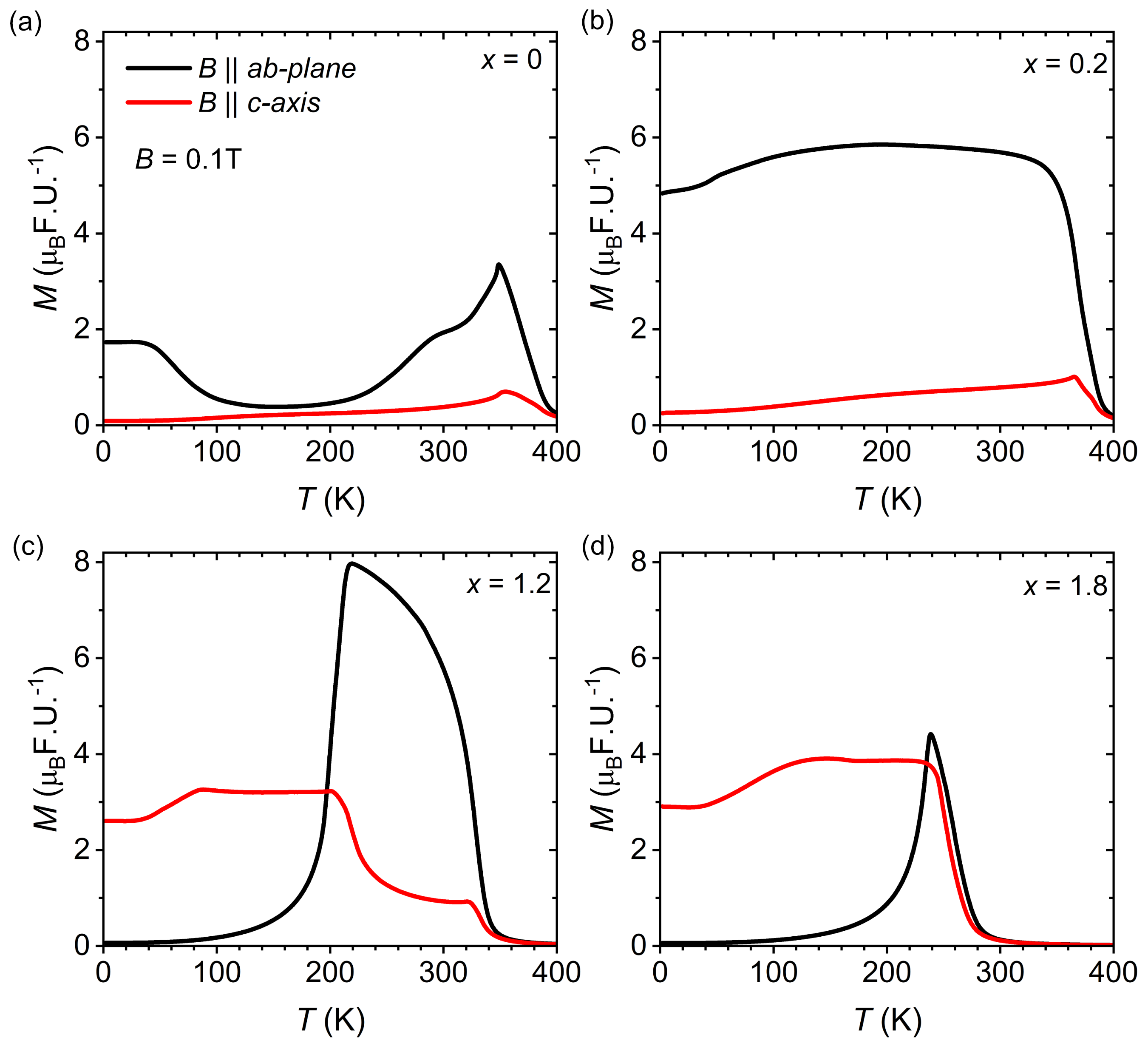}
    \caption{Temperature dependent magnetization (normalized by formula unit) measurements of TmMn$_6$Sn$_{6-x}$Ga$_x$ with applied field of 0.1 T, along the $ab$ plane (black) and along the $c$-axis (red). (a) Magnetization curves for the parent compound ($x = 0$), (b) Ga concentration of $x = 0.2$, (c) Ga concentration of $x = 1.2$, and (d) Ga concentration of $x = 1.8$. The data illustrate the transition from easy-plane anisotropy in low Ga concentrations to easy-axis anisotropy with higher Ga concentrations, highlighting the spin reorientation.}
    \label{Fig4}
    \end{center}
\end{figure}

\begin{figure*}[!ht]
\begin{center}
\includegraphics[scale=0.12]{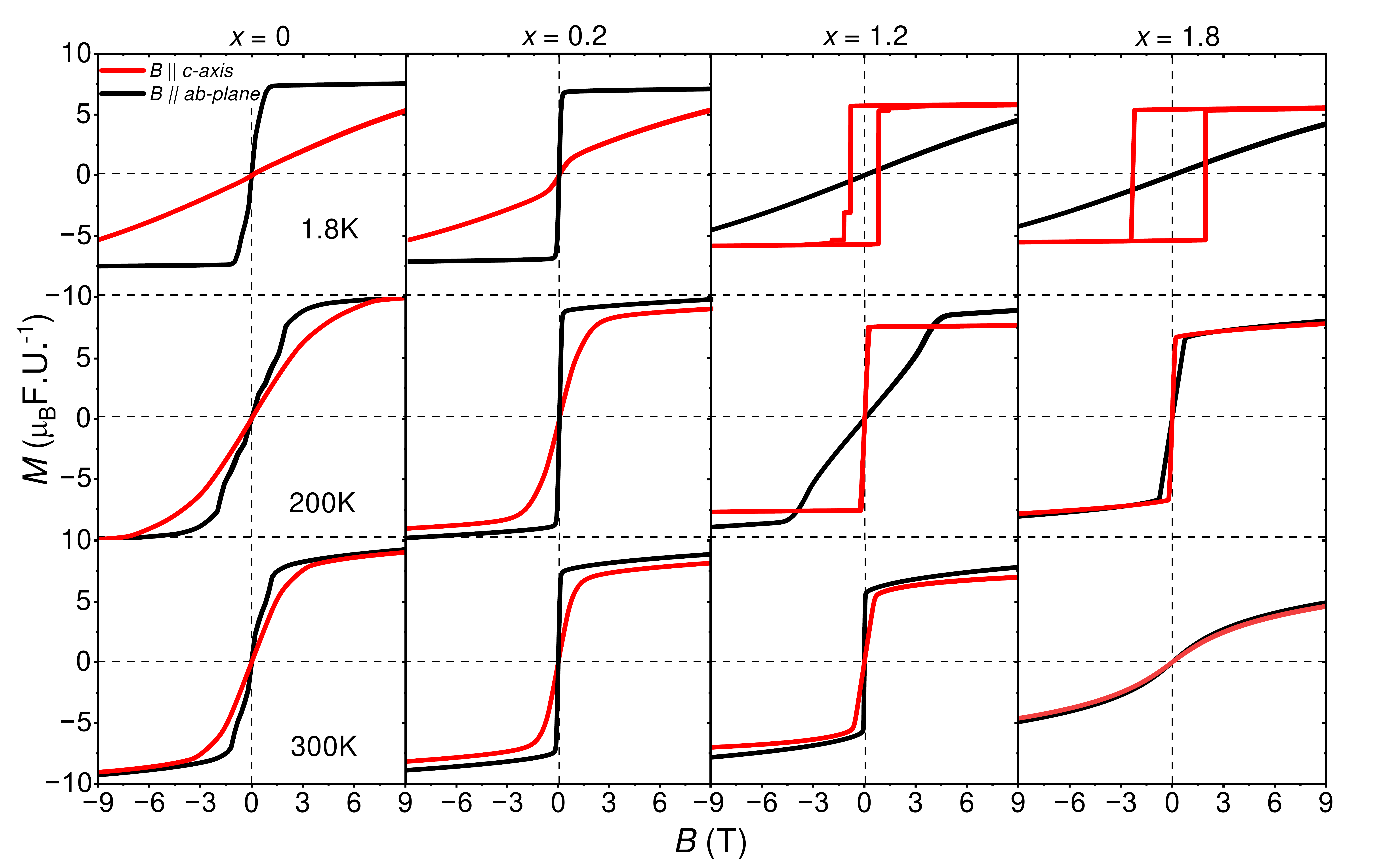}
    \caption{Magnetic field-dependent magnetization of TmMn$_6$Sn$_{6-x}$Ga$_x$ at selected temperatures (1.8 K, 200 K, 300 K). (a) $x = 0$, (b) $x = 0.2$, (c) $x = 1.2$, and (d) $x = 1.8$. The plots show saturation behavior that changes with temperature and Ga concentration, providing further evidence for the spin reorientation transition.}
    \label{MH}
    \end{center}
\end{figure*}

The magnetic behavior of TmMn$_6$Sn$_{6-x}$Ga$_x$ evolves significantly with Ga doping, as shown by thermomagnetic measurements performed under a 0.1 T magnetic field applied along the $c$-axis ($M_{c}$, red) and within the $ab$-plane ($M_{ab}$, black), presented in Fig. \ref{Fig4}.  For the undoped compound ($x=0$) (Fig. \ref{Fig4}(a)), a clear peak at 348 K marks the transition to the  collinear antiferromagnetic state. With further decrease in temperature, a cusp-like feature appears in $M_{ab}$ around 330 K corresponding to the antiferromagnetic to helimagnetic transition, consistent with prior neutron diffraction studies \cite{lefevre_neutron_2002}. Additional features are observed near 300 K and 50 K in the $ab$-plane response only. Notably, across the entire temperature range, $M_{ab}$ remains consistently larger than $M_{c}$, consistent with previously determined \cite{lefevre_neutron_2002} easy-plane anisotropy of the parent compound TmMn$_6$Sn$_6$. 

Upon low Ga doping ($x = 0.2$, Fig.\ref{Fig4}(b)), we observe a marked change in the magnetic response, particularly along the $ab$-plane, while $M_{c}$ remains largely unaffected. The observed behavior is typical of a ferrimagnet (FiM), in agreement with prior neutron diffraction studies \cite{lefevre_neutron_2002}. Once again, the highlight of this measurement is that the net moment in the $ab$-plane remains dominant over that along the $c$-axis, confirming the persistence of easy-plane anisotropy at this doping level.

\begin{figure}
  \begin{center}
  \includegraphics[scale=0.041]{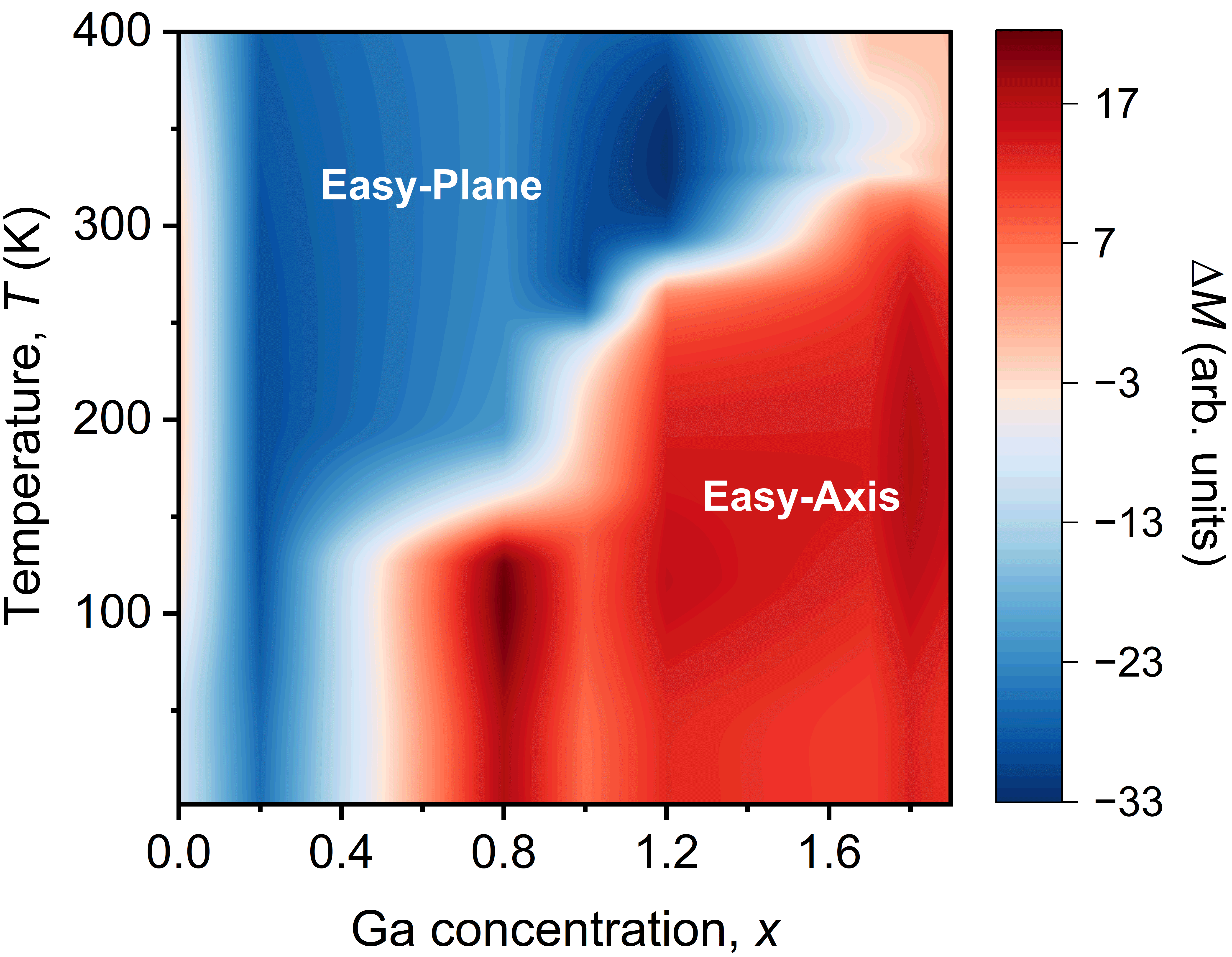}
  \caption{Magnetic phase diagram for TmMn$_6$Sn$_6$–xGa$_x$ as a function of Ga concentration and temperature under a magnetic field of 0.1 T. The blue regions indicate easy-plane anisotropy, while the red regions correspond to easy-axis anisotropy. The white region, present only in samples undergoing spin reorientation, marks the boundary where the competing magnetic anisotropies are maximal.}
  \label{PD}
  \end{center}
\end{figure}

However, as the Ga concentration increases to $x = 1.2$, the magnetic behavior dramatically changes. Around 200 K, $M_{ab}$ (black curve) drops sharply while $M_{c}$ (red curve) increases. Below this temperature, $M_{c}$ dominates, indicating easy-axis anisotropy; above it, the system retains easy-plane character. This marks a clear spin reorientation transition from easy-plane to easy-axis anisotropy upon cooling. Similarly, the $x = 1.8$ (TmMn$_6$Sn$_{4.2}$Ga$_{1.8}$) composition in Fig. \ref{Fig4}(d) displays a spin reorientation at $~240$ K, where the easy-plane anisotropy temperature range is rather small, but not zero.

Field-dependent magnetization measurements (Fig. \ref{MH}) at selected temperatures (1.8 K, 200 K, and 300 K) further support the presence of a spin reorientation. For $x = 0$ and $x = 0.2$, the low-temperature $M(B)$ curves reflect easy-plane anisotropy, with $M_{ab}$ saturating more rapidly than $M_c$. At higher temperatures, the difference between in-plane and out-of-plane responses diminishes, indicating a trend towards maximally competing anisotropies.
In contrast, for $x = 1.2$ and $x = 1.8$, the curves reflect easy-axis anisotropy, particularly at low temperatures, with $M_c$ overtaking $M_{ab}$. For $x = 1.2$, the intermediate temperature behavior (e.g., 200 K) shows a crossover, signaling the transition from easy-plane to easy-axis anisotropy across the spin reorientation boundary.

The evolution of magnetic anisotropy as a function of Ga content and temperature is summarized in the phase diagram shown in Fig. \ref{PD}. The phase diagram, constructed by subtracting the $c$-axis magnetic moment from the $ab$-plane magnetic moment as a function of both concentration and temperature under a field of 0.1 T, provides a conclusive pattern for the magnetic behavior of the full doping series we grew (Fig. \ref{PD}). Easy-plane and easy-axis regimes are marked in blue and red, respectively, while the white region highlights the region of maximally competing anisotropies associated with the spin reorientation transition seen above a critical Ga doping level ($x=0.5$). The boundary of this region marks the onset of the spin reorientation.

\subsection{Electrical Transport Properties}

\begin{figure}[!ht]
\begin{center}
\includegraphics[scale = 1]{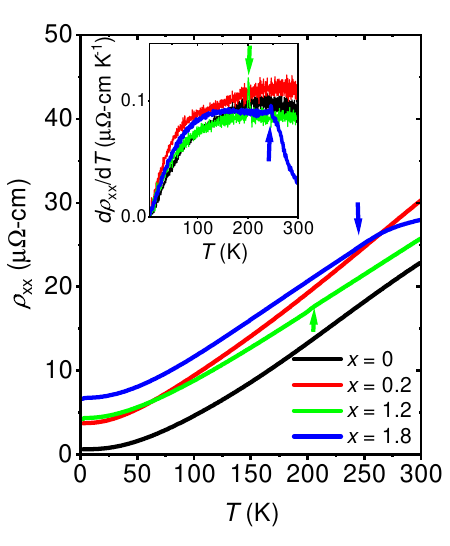}
    \caption{The electrical resistivity ($\rho_{xx}$) of four compositions of TmMn$_6$Sn$_{6-x}$Ga$_{x}$, measured with current applied in the $ab$-plane, as a function of temperature. The inset shows the temperature derivative of the resistivity. Arrows serve as guides to the eye, highlighting features associated with spin reorientation in the $x = 1.2$ and $1.8$ compositions, visible in both the resistivity and its derivative.} 
    \label{Rho}
    \end{center}
\end{figure}

The temperature-dependent electrical resistivity $\rho_{xx}$, measured between 1.8 K to 300 K with current applied along a random direction within the $ab$-plane, is shown in Fig. \ref{Rho}, for four compositions: $x= 0, 0.2, 1.2,$ and $1.8$. All samples exhibit metallic behavior. As expected, the residual resistivity increases significantly from $x = 0$ to $x = 0.2$, primarily due to doping-induced disorder, while further increase in $x$ lead to only modest additional changes. For $x = 0$ (black) and $0.2$ (red), the resistivity curves are nearly parallel over the entire temperature range, suggesting similar scattering mechanisms. In contrast, the compositions exhibiting spin reorientation ($x = 1.2$ and $1.8$) display a change in the resistivity slope, likely due to variations in the spin-dependent scattering rate associated with the spin reorientation. A subtle kink appears at the spin reorientation temperature for both $x = 1.2$ and $1.8$, as marked by arrows in Fig. \ref{Rho}. These anomalies become more apparent in the temperature derivative of the resistivity, shown in the inset of Fig. \ref{Rho}.

\begin{figure}
\begin{center}
\includegraphics[scale =1.1]{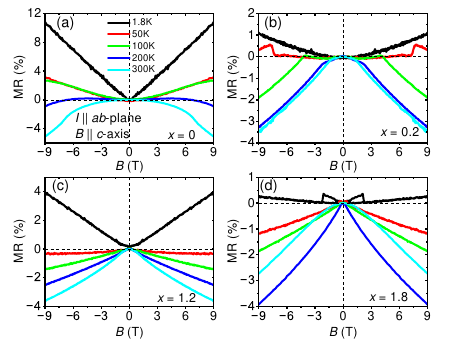}
    \caption{Magnetoresistance (MR) as a function of magnetic field ($B$) applied along the $c$-axis, with current ($I$) applied in the $ab$-plane, measured at 1.8, 50, 100, 200, and 300 K for four compositions of TmMn$_6$Sn$_{6-x}$Ga$_{x}$: (a) undoped parent compound ($x = 0.0$), (b) lightly doped ($x = 0.2$), (c) intermediate doping ($x = 1.2$), (d) heavily doped ($x = 1.8$). The evolution of the magnetoresistance with doping and temperature reveals changes in spin scattering and underlying magnetic anisotropy, particularly near the spin reorientation regime.} 
    \label{MR}
    \end{center}
\end{figure}

Transverse magnetoresistance (MR) measured with current in the $ab$-plane and magnetic field along the $c$-axis, defined by MR = [($\rho_{xx}({B}$-$\rho_{xx}({B=0})$/$\rho_{xx}({B=0})$)]$\times$100 \% for the four compositions is shown in Fig. \ref{MR}. The MR practically tracks the magnetization curves depicted in Fig. \ref{MH} in all four compositions, the bow-tie MR is clearly visible in $x=1.8$ composition at 1.8 K and 50 K, marking the hysteretic behavior observed in $M$ vs $B$. This feature, however, is not as clear in $x=1.2$ composition. The parent compound shows a linear MR behavior at 1.8 K (see Fig. \ref{MR}(a)), similar to that in Y166, before the Lifshitz transition \cite{siegfried_magnetization-driven_2022}.

\begin{figure}[!ht]
\begin{center}
\includegraphics[scale = 1.1]{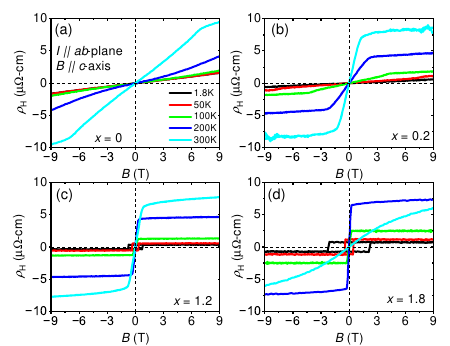}
    \caption{Hall resistivity $\rho_{\text{H}}$ as a function of magnetic field ($B$) applied along the $c$-axis, with current ($I$) applied in the $ab$-plane, measured at 1.8, 50, 100, 200, and 300 K for four compositions of TmMn$_6$Sn$_{6-x}$Ga$_{x}$: (a) parent compound TmMn$_6$Sn$_6$ ($x = 0.0$), (b) lightly doped $x = 0.2$, (c) intermediate doping $x = 1.2$, (d) heavily doped $x = 1.8$. The evolution of the Hall response reflects doping-induced modifications to the magnetic structure and spin reorientation behavior across the series.} 
    \label{Hall}
    \end{center}
\end{figure}

The Hall resistivity, $\rho_{\text{H}}$, measured with current applied in the $ab$-plane and magnetic field along the $c$-axis, is presented in Fig. \ref{Hall}. The field dependence of $\rho_{\text{H}}$ closely mirrors the magnetization curves discussed in Fig. \ref{MH}. In particular, $\rho_{\text{H}}$ exhibits a trend that tracks the net magnetization $M$ across all compositions and temperatures, as expected for ferrimagnetic systems as similarly observed in other $R$166 compounds \cite{siegfried_magnetization-driven_2022,bhandari_magnetism_2024,bhandari2024three,kabir2022unusual}.

\subsection{Heat Capacity}

\begin{figure}[!ht]
\begin{center}
\includegraphics[scale = 1.7 ]{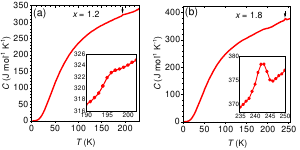}
    \caption{Temperature dependence of the heat capacity $C_V$ for TmMn$_6$Sn$_{6-x}$Ga$_{x}$ at doping levels (a) $x = 1.2$ and (b) $x = 1.8$. A subtle kink/hump is observed in both compositions—near $~196 K$ for $x = 1.2$ and $~243 K$ for $x = 1.8$—highlighting  the spin reorientation transition.} 
    \label{Fig9}
    \end{center}
\end{figure}


At the spin reorientation temperature, the anisotropies of the Mn and rare-earth sublattices become comparable, and the system transitions between two competing anisotropic directions—corresponding to distinct local minima in the magnetic free energy landscape. This reorientation involves a collective realignment of spins and a shift in the direction of the magnetization vector, $\mathbf{M}$. Because this realignment changes produces a thermodynamic anomaly—observable as a kink or step in the heat capacity.

The thermodynamic signature of the engineered spin reorientation in TmMn$_6$Sn$_{6-x}$Ga$_{x}$ is clearly captured in the heat capacity measurements shown in Fig. \ref{Fig9}. In particular, the Ga-doped compositions ($x = 1.2$ and $x = 1.8$) exhibit a pronounced anomaly—manifesting as a kink—in the heat capacity curves, centered around the spin reorientation temperature (196 and 241 K, respectively). This feature reflects the second-order magnetic transition associated with the gradual reconfiguration of spin orientation from in-plane to out-of-plane. The emergence and shifting of this feature with Ga content further confirms the tunability of magnetic anisotropy through chemical substitution, providing a direct thermodynamic validation of the spin reorientation transition.

\subsection{First-Principles Calculations}

A previous study has confirmed that the Mn and rare-earth $4f$ moments are governed by very different spin dynamics as the latter experience much more rapid fluctuations \cite{jones_origin_2022}. 
At low temperatures, the magnetic anisotropy (MA) in these compounds is predominantly determined by the $4f$ contribution due to their strong spin-orbit coupling. 
As the temperature increases, the magnetic moments of the rare-earth elements, such as Tm, decline at a faster rate compared to the Mn moments. 
Therefore, the contribution of the $4f$ moments to the MA becomes less significant at higher temperatures.

\begin{figure}
  \begin{center}
  \includegraphics[scale=0.49]{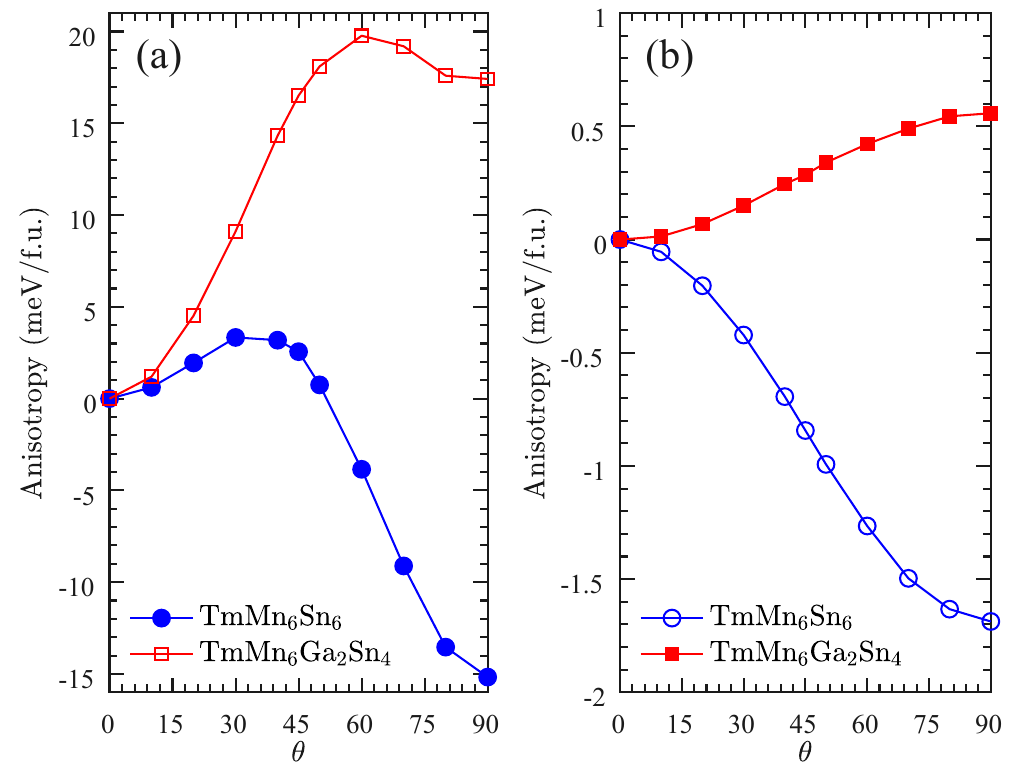}
  \caption{Magnetic energy variation (meV/f.u.) plotted against spin-axis rotation
for TmMn$_{6}$Sn$_{6}$ and TmMn$_{6}$Sn$_{4}$Ga$_{2}$ with (a)
and without (b) Tm-$4f$ contributions. Note the difference in scale in panels (a) and (b).}
  \label{tm166ga}
  \end{center}
\end{figure}

We begin our analysis by looking at the overall magnetic anisotropy which corresponds to the low-temperature condition where the $4f$-electron contribution to the MA dominates. Figure \ref{tm166ga} (a) displays the total energies $E(\theta)$ obtained from WIEN2k DFT+$U$ calculations, plotted as functions of the direction {of the magnetic moments}, characterized by the angle $\theta$ with respect to the $c$ axis, for TmMn$_{6}$Sn$_{6}$ and TmMn$_{6}$Sn$_{4}$Ga$_{2}$. The results reveal that the easy-plane anisotropy in TmMn$_{6}$Sn$_{6}$ transitions to the easy-axis anisotropy in TmMn$_{6}$Sn$_{4}$Ga$_{2}$, consistent with low-temperature measurements. The non-monotonic dependence suggests that the presence of large higher-order anisotropy constants in the $R166$ system, which, as previously shown \cite{lee2023prb}, is due to the fact that the SOC is much larger than the crystal-field (CF) splitting. The substantial change in calculated anisotropy implies a significant change in the single-ion Tm-$4f$ anisotropy due to Ga-doping-induced CF changes. 
In our earlier study, we confirmed this by treating the CF as a perturbation and calculating the variation of CF energy when the spin quantization axis (i.e., the magnetic moment direction) is rotated from $\hat{z}$ to the basal plane~\cite{lee2023prb}; we were able to qualitatively 
reproduce the results in Fig. ~\ref{tm166ga} (a). 

In order to separate Tm and Mn contributions, 
we performed open-core calculations that treated occupied Tm $4f$-electrons
as core electrons without spin-polarization (albeit Tm $d$-electrons are still included --- but their contribution is likely small). The results are presented
in Fig. ~\ref{tm166ga} (b) (WIEN2k calculations) and Table \ref{tab:j_anis} (VASP calculations). 
In Fig. ~\ref{tm166ga} we present the $total$ magnetic anisotropy energy (MAE), including single site anisotropy ($A$ in Eq. \ref{eq:spin_ham}), as well as anisotropic exchange components $B_1$ and $B_2$. In Table \ref{tab:j_anis}  we present them separately. Note that the difference between the total energy at $\theta=0$ and $\theta=90^\circ$ in Fig. ~\ref{tm166ga} should be compared to $2A+B_1+B_2$ in 
Table \ref{tab:j_anis}; indeed the numbers are quite similar, albeit VASP systematically overestimates the anisotropy parameters. The most important message is that in most cases (as pointed out before \cite{ghimire_competing_2020}) the single site anisotropy is actually easy-axis, and the net easy-plane anisotropy of Mn planes is determined by the anisotropic exchange --- and while the latter only change so much between the systems, the single site anisotropy is rather sensitive to chemical composition of the rare-earth/(Sn,Ga) level.

As discussed above, for a reorientation transition one has to have opposite anisotropies for the rare earth (controlling the low temperatures) and Mn (controlling the high temperatures). This does hold for Tb166 \cite{ghimire_competing_2020}, but the calculations in Fig. ~\ref{tm166ga} predict that there will be no reorientation at $x=0$ (only easy plane), nor at $x=2$ (only easy axis). Interpolating between there two high-symmetry calculations, we conclude that the reorientation transition must appear at some small $x$ at a small temperature, the transition temperature should grow continuously with $x$, and at some $x\lesssim 2$ merge with the Curie temperature. This conclusion is in good qualitative agreement with the experiment (Fig. \ref{PD}).
\begin{table}
\begin{tabular}{r@{\extracolsep{0pt}.}lr@{\extracolsep{0pt}.}lr@{\extracolsep{0pt}.}lr@{\extracolsep{0pt}.}lr@{\extracolsep{0pt}.}l}
\multicolumn{2}{c}{} & \multicolumn{4}{c}{TmMn$_{6}$Sn$_{6-x}$Ga$_{x}$} & \multicolumn{4}{c}{YMn$_{6}$Sn$_{6-x}$Ga$_{x}$}\tabularnewline
\hline 
\multicolumn{2}{c}{$x$} & \multicolumn{2}{c}{0} & \multicolumn{2}{c}{2} & \multicolumn{2}{c}{0} & \multicolumn{2}{c}{2}\tabularnewline
\hline 
\multicolumn{2}{c}{} & \multicolumn{8}{c}{(eV)}\tabularnewline
\hline 
\multicolumn{2}{c}{$J_{1}$} & -0&0691 & -0&0594 & -0&0716 & -0&0575\tabularnewline
\multicolumn{2}{c}{$J_{2}$} &  0&0140 & -0&0838 &  0&0108 & -0&0599\tabularnewline
\hline 
\multicolumn{2}{c}{} & \multicolumn{8}{c}{(meV)}\tabularnewline
\hline 
\multicolumn{2}{c}{$A$}     & 0&200 & -0&732 & -0&112 & -0&640\tabularnewline
\multicolumn{2}{c}{$B_{1}$} & 1&138 &  0&484 &  1&243 &  0&836\tabularnewline
\multicolumn{2}{c}{$B_{2}$} & 0&683 &  0&174 &  0&998 &  0&218\tabularnewline
\hline 
\end{tabular}
\caption{\label{tab:j_anis}Exchange coupling parameters and single-site anisotropy {(per plane)} 
for RMn$_{6}$Sn$_{6-x}$Ga$_{x}$ where R=Y and Tm, and $x=0$ and
$2$. }
\end{table}

Finally, let us quickly discuss the helimagnetic phase that appear only at very small dopings $x\approx 0$. As we know, the prerequisit for this phase is that the three nearest neighbor exchange couplings, $J_1$, $J_2$ and $J_3$ are frustrated. In the parent Y compound the $J_1$ and $J_3$ interactions are ferromagnetic, and $J_2$ is antiferromagnetic, thus, according to the mean field theory in Ref. \cite{ROSENFELD20081898}, it features a double-spiral with the two rotation angles of the opposite signs ($H2$, in their notation). There is no reason to expect $J_1$ to depend on either the rare earth or Ga doping, and, indeed, it remains ferromagnetic in all four cases in Table \ref{tab:j_anis}. $J_2$ changes sign upon Ga doping in Y166 (not probed experimentally yet), so we predict that a collinear state will form upon such doping. In Tm compounds, the Mn-Mn $J_2$ shows the same effect. Since Mn-Tm exchange also contributes to $J_2$, it will largely offset the positive value calculated in Table \ref{tab:j_anis}, so that the absolute value will be small, but either sign is possible. This present an interesting case within the theory of Ref. \cite{ROSENFELD20081898}: if $J_2$ is small, but positive, while $J_3$ is negative, we can get the same $H2$ spiral, but if $J_2<0$ we can get either ferromagnetism, if $J_3<0$, or a $H1$ spiral, where interplane rotations are always in the same direction. Similarly, if $J_2>0$ we can get, besides the $H2$ spiral as in Y166, only a collinear AF state. 

A previous study based on neutron diffraction has shown that TmMn$_{6}$Sn$_{6}$,  hosts an $H1$ helimagnetic ordering over a large temperature range from $T=2$ K to $T=330$ K and transitions into a collinear AF state only slightly below $T_{N}=347$ K \cite{lefevre2003neutron}.
In view of the above analysis, and using both the experiment and the DFT calculations, we can explain the full magnetic phase diagram thus:

1. $J_1<0$ is always FM; at very small doping $|J_2|\ll |J_1|$, and below 330 K $J_2<0$, while above $J_2>0$. At all temperatures, $J_3>0$. In this entire range, both Tm and Mn have easy-plane anisotropy.

2. As doping increases, $J_3$ rapidly changes sign, and the ground state becomes in-plane FM, at all temperatures.

3. When doping further increases to $x=0.5$--0.6, the anisotropy of Tm changes from easy-plane to easy axis, with the two being nearly degenerate states, and intermediate directions strongly unfavorable. At this point, a reorientation transition appears, at low temperatures.

4. Even further doping leads to a rapid growth in Tm anisotropy, and reduction of the Mn one, thus the reorientation temperature rapidly grows. When $x$ reaches $\sim 2$, this temperature merges with the N\`eel temperature, and no reorientation is observed any further. At all temperatures the system remains an easy-axis ferrimagnet.

\subsection{Concluding Remarks}

$R$Mn$_6$Sn$_6$ compounds exhibit a fascinating crystal structure that allows the stabilization of various magnetic states depending on the choice of the $R$ atom. In these systems, Mn exhibits in-plane magnetic anisotropy, while the $R$ atoms favor an out-of-plane (easy-axis) magnetic orientation. In TbMn$_6$Sn$_6$ (Tb166), the anisotropy of Tb dominates over that of Mn at 310 K, leading to a spin-reorientation transition in which the magnetic moments reorient from the $ab$-plane to the $c$-axis. In contrast, the moments in TmMn$_6$Sn$_6$ (Tm166) remain in the $ab$-plane, yet the system exhibits a rich magnetic phase diagram: it first orders into a collinear antiferromagnetic state below 350 K and then transitions into a spiral state below approximately 330 K.

A small amount of Ga intercalation renders the system ferrimagnetic. With increasing Ga concentration, the easy-axis anisotropy becomes more pronounced, and the system undergoes a temperature-dependent spin-reorientation. The spin-reorientation temperature increases systematically with Ga content. Our first-principles calculations, consistent with experimental observations, show that at and above $x = 2$, the compound exhibits overall easy-axis anisotropy, whereas at $x = 0$, in-plane anisotropy dominates. At intermediate Ga concentrations, the spin reorientation arises from the interplay between the magnetic anisotropies of the Tm and Mn atoms, modulated by Ga intercalation.

Magnetic biskyrmion textures have been observed near the spin-reorientation transition in TbMn$_6$Sn$_6$ \cite{li_discovery_2023}. Similarly, we have observed skyrmion bubbles just below the spin-reorientation transition temperature in TmMn$_6$Sn$_{4.2}$Ga$_{1.8}$ \cite{gazzah2025skyrmion}. In this study, we demonstrate that the spin reorientation persists in the range $x = 1.2$ to $1.8$, with the reorientation temperature increasing from ~196 K to ~241 K. It would be intriguing to carry out a finely tuned doping study to investigate its effect on the formation, stability, and size of skyrmionic textures.

Moreover, Ga doping is known to modify the spin-reorientation transition in Tb166 \cite{perry2006anisotropic}, in which—unlike in Tm166—the transition occurs spontaneously without doping. This suggests that similar spin reorientations can be induced in other $R$Mn$_6$Sn$_6$ compounds, particularly those with magnetic $R$ elements. This opens a broad materials space for studying the interplay between exchange interactions ($J$ values) and magnetic anisotropy—factors that are critical for stabilizing skyrmionic textures. Such studies may pave the way for achieving smaller, tunable skyrmions with large topological Hall signals as predicted in Ref. \cite{wang2020skyrmion}.



Another noteworthy study on the spin reorientation phenomena is shown in TbMn$_{6}$Sn$_{6}$ by Ryan et al. \cite{ryan2024}, where they have determined the timescale of the spin reorientation to be of about 20 ns. Since a spin reorientation is governed by the strength of the exchange interactions, tuning the spin reorientation temperature in TmMn$_{6}$Sn$_{6}$ provides a means to investigate and control spin dynamic - an area of significant interest in spintronics. This study, therefore, not only sheds light on TmMn$_{6}$Sn$_{6}$ but also suggests broader investigations across the RMn$_6$Sn$_6$ family using the diverse dopant types and concentrations.

\begin{acknowledgments}
N.J.G acknowledges the support from the NSF CAREER award DMR-2143903. I.I.M. acknowledges support from the National Science Foundation under Award No. DMR-2403804.
Work at Ames National Laboratory is supported by the U.S.~Department of Energy, Office of Basic Energy Sciences, Materials Sciences and Engineering Division.
LK acknowledges Research Computing at the University of Virginia for providing computational resources for a part of this research.
Ames National Laboratory is operated for the U.S.~Department of Energy by Iowa State University under Contract No.~DE-AC02-07CH11358. Work at Argonne National Laboratory (a set of EDS measurements) was supported by the U.S. Department of Energy, Office of Science, Basic Energy Sciences, Materials Science and Engineering Division.
\end{acknowledgments}

\end{document}